# The Design of Compressive Sensing Filter

Lianlin Li, Wenji Zhang, Yin Xiang and Fang Li

Institute of Electronics, Chinese Academy of Sciences, Beijing, 100190

Lianlinli1980@gmail.com

*Abstract:*

**In this paper, the design of *universal* compressive sensing filter based on *normal* filters including the lowpass, highpass, bandpass, and bandstop filters with different cutoff frequencies (or bandwidth) has been developed to enable signal acquisition with sub-Nyquist sampling. Moreover, to control flexibly the size and the coherence of the compressive sensing filter, as an example, the microstrip filter based on defected ground structure (DGS) has been employed to realize the compressive sensing filter. Of course, the compressive sensing filter also can be constructed along the identical idea by many other structures, for example, the man-made electromagnetic materials, the plasma with different electron density, and so on. By the proposed architecture, the *n*-dimensional signals of *S*-sparse in *arbitrary* orthogonal frame can be exactly reconstructed with measurements on the order of *S*log($n$) with overwhelming probability, which is consistent with the bonds estimated by theoretical analysis.**

I.  INTRODUCTION

Advances in computation power have enabled digital signal processing to become the primary modality in many applications, such as, communications, multimedia, and radar detection systems. Converting analogy signals to the digital ones avoids the complicated design considerations for analog processing. The theoretical base of the traditional ADCs, such as flash ADCs, pipelined ADCs and sigma-delta ADCs, is the so-called Nyquist-Shannon theorem which guarantees the reconstruction of a band-limited signal when it is uniformly sampled with a rate of at least twice its bandwidth. Consequently, the physical limitation of traditional analogy-to-digital converters is the main obstacle towards pushing their performance to the GHz-regime. As we known, the uniform sampling is not a very efficient technique in extracting the information out of sparse signals because of only the *prior* information, the signal bandwidth or approximate bandwidth is used for the signal sampling based on the Nyquist-Shannon theorem. However, many signals of interest have additional *structure*, which can be called sparsity or compressibility. Recently, a new emerging field has made a paradigmatic step in the way information is presented, stored, transmitted and recovered. This area is often referred to as compressive sensing (or compressed sensing, compressed sampling, etc) developed by Donoho, Tao, Candes and Romberg et al [1-4]. The CS theory asserts that one can recover certain signal or image from far fewer samples or measurements than traditional methods required when the signal of interest is compressible or sparse in some basis.

The CS measurements, different than samples that traditional analogy-to-digital converters take, model the acquisition of signal $x_0$ as a series of inner products against different the independent waveforms $\{\phi_k : k = 1, 2, 3, \cdots, m\}$, in particular,

$$y_k = \langle \phi_k, x_0 \rangle, \quad k = 1, 2, 3, \cdots, m \tag{1.1}$$

As well known, the recovering $x_0$ from $y_k$, a kind of classical linear inverse problem will need more measurements than unknowns, i.e. $m \geq n$. But the CS theory tell us that if the signal of interest $x_0$ is $S$-sparse in the orthogonal framework of $\Psi$ and the $\phi_k$ are chosen appropriately, then results from CS have shown us that recovering $x_0$ is possible even when there are far fewer measurements than unknowns, $m \ll n$. We say $x_0$ is S-sparse in $\Psi$ if we can decompose $x_0$ as $x_0 = \Psi \alpha_0$, where $\alpha_0$ has at most $S$ non-zero components. In some applications, the signals of interest are not perfectly sparse; however, all most of information can be captured by small number of terms. That is, there is a transform vector $\alpha_{0,S}$ with only S terms such that $\|\alpha_{0,S} - \alpha_0\|_2$ is small. Given the measurements $y = \Phi x_0$, we solve the convex optimization

program

$$\min_{\alpha} \|a\|_{l_1} \quad \text{subject to} \quad y = \Phi\Psi\alpha \tag{1.2}$$

The $l_1$-norm is being used to measure the sparsity of candidate signals.

By considering recovery stochastically, it has been shown that measurements generated from Gaussian or Bernoulli random variables can ensure the signal recovery with high probability. Following this theory, the well-known single pixel camera has been invented by Baraniuk et al. Later, many efforts to design the *universal* CS measurement instruments have been done by many authors, for example, the chip-level Analogy-to-Information converter by Ragheb, Baraniuk et al [11], the single-shot compressive spectral imager proposed by M. E. Gehm et al [12], and so on. But, these CS measurements can not be usually used in practice (at least can not used for the real-time purpose) because of its time-consuming computation and the difficulty of physical realization. Vertterli et al has developed alternative approach named as sampling signal with finite rate of innovation. The center idea is that the sampling rate for a sparse signal can be significantly reduced by first convolving with a kernel that spread it out. In [8], the numerical simulations are carried out to demonstrate the recovery of sparse signals from a small number of samples of the output of a finite length "random filter". In [9] Romberg has developed a *universal* CS measurement by using a special random convolution (its amplitude of frequency-domain identically equal to 1) and derived bounds on the number of samples need to guarantee sparse reconstruction from a more theoretical perspective. The center idea of Romberg's filter can be summarized as: through the CS filter, the signal frequency-domain phase is modulated by random waveform while the signal frequency-domain amplitude is not distorted; consequently, the resulting signal will be spread out in time domain. Along the Romberg's idea, L. Jacques et al have constructed the CMOS compressed imaging by a shift register set in a pseudo-random configuration.

## II. THE DESIGN OF COMPRESSIVE SENSING FILTER

In this paper, the novel design of compressive sensing filter has been proposed by using the normal microwave filters, the plasma with different electron density, and so on. To modulate the signal frequency-domain phase by random waveform while the amplitude is kept, the signal components corresponding to different frequencies should be extracted from the time-domain signal and be modulated. Obviously, a series of filters with different cutoff frequencies followed by transmission line with random length by which the random phase modulation can be readily realized will be good candidates. Following this, we design the compressive sensing filter. Refer to Fig.1 for the example where the normal lowpass filters with different cutoff frequencies are employed; however, the lowpass, bandstop and bandpass filters also can be used. It should be pointed out that:

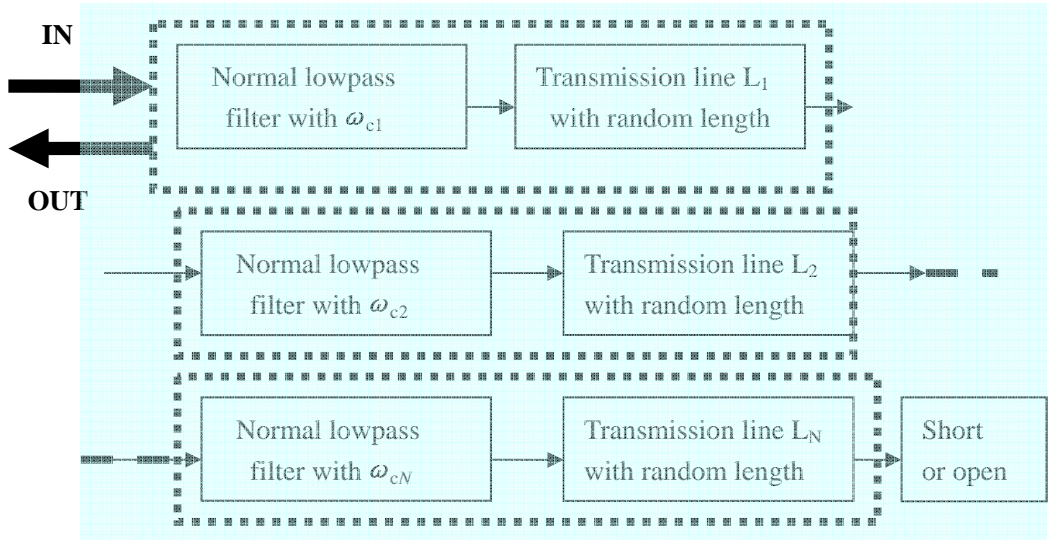

Fig.1. the scheme map of the proposed compressive sensing filter by the lowpass filters

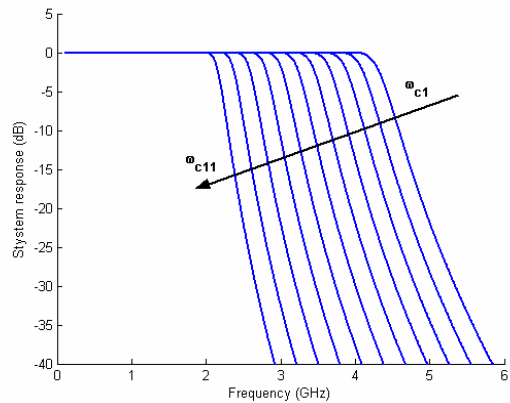

Fig.2  System responses of eleven ideal 9-order Chebyeshev lowpass

filters for different cutoff frequencies

(1) $\omega_{c1} > \omega_{c2} > \cdots > \omega_{cN}$ for Fig.1, where $\omega_{ck}$ is the cutoff radian frequency of the $k$th lowpass filter. Taking an example, the system responses of eleven 9-order Chebyshev highpass filters is shown in Fig.2. Consequently, the signal components with frequencies smaller than $\omega_{c1}$ will be

firstly received, and then the signal components with frequencies smaller than $\omega_{c2}$ but bigger than $\omega_{c1}$ which be modulated by the transmission line $L_1$ with random length will be received, and so on. Obviously, the signal in time domain has been spread out!

(2) to realize the random coded phase with respect to different frequencies, the transmission line with random length has been employed; Of course, many microwave elements can be employed to control flexibly the random phase, for example, the PIN diode.

(3) to keep the frequency-domain amplitude of signal, the short or open at the ended port is specified.

Moreover, it is noted that the proposed compressive sensing filter belongs to the single-port net, in order to measurement the signal, the circle-instrument or coupler should be used in practice. To decrease the size of proposed compressive sensing filter, the microstrip filter based on the well-known defected ground structure (DGS) proposed by J. I. Park et al [13] has been employed to realize the unit of compressive sensing filter (see Fig.3 for a unit, whose cutoff frequency is 3GHz and size of ~3cm).

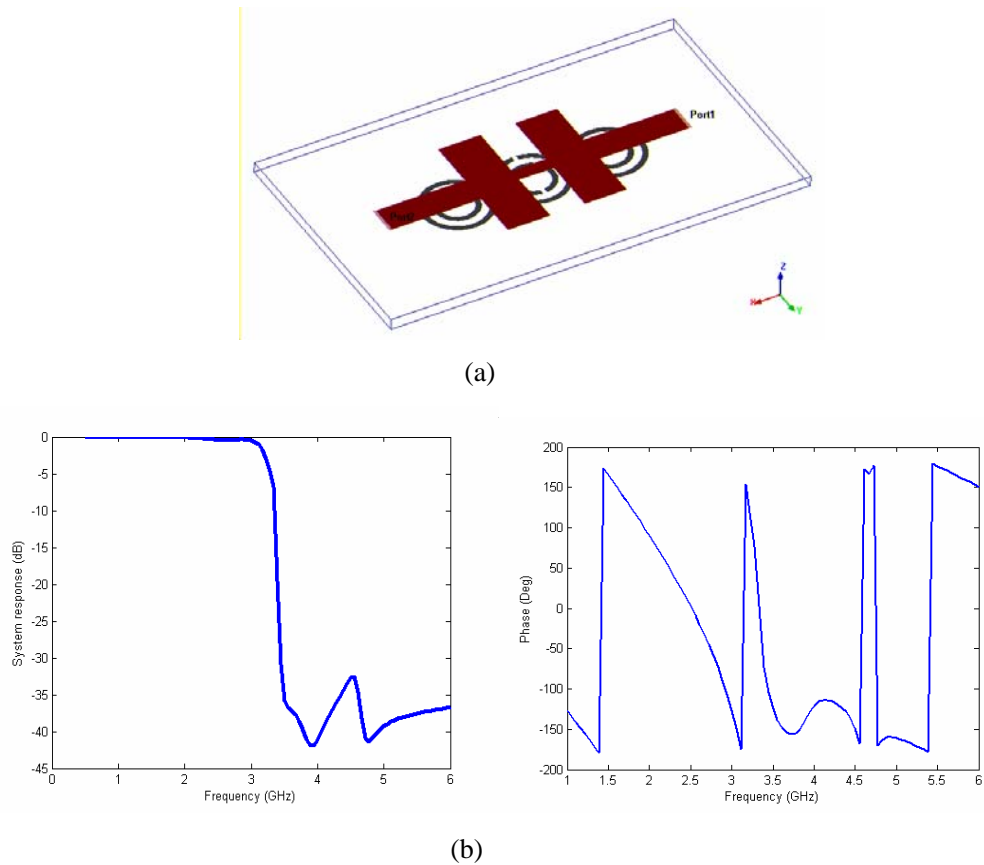

(a)

(b)

Fig.3 (a) the DGS unit of proposed compressive sensing filter and (b) its system response

## III. SYSTEM SIMULATIONS

In this section several simulations are provided to verify the proposed *universal* compressive sensing filter. The compressive sensing filter consist of 11 DGS microstrip filters whose cutoff frequency from 2GHz to 4GHz with separation 0.2GHz. The signal is sampled at 1/5 Nyquist rate and is reconstructed by so-called Bayesian compressive sensing method. As a first example, Fig.4(a) shows the reconstruction of 26-sparse sparse signal in time domain and Fig.4(b) shows the sampled signal. As a second example, the reconstruction of 26-sparse signal in frequency domain is shown in Fig.5.

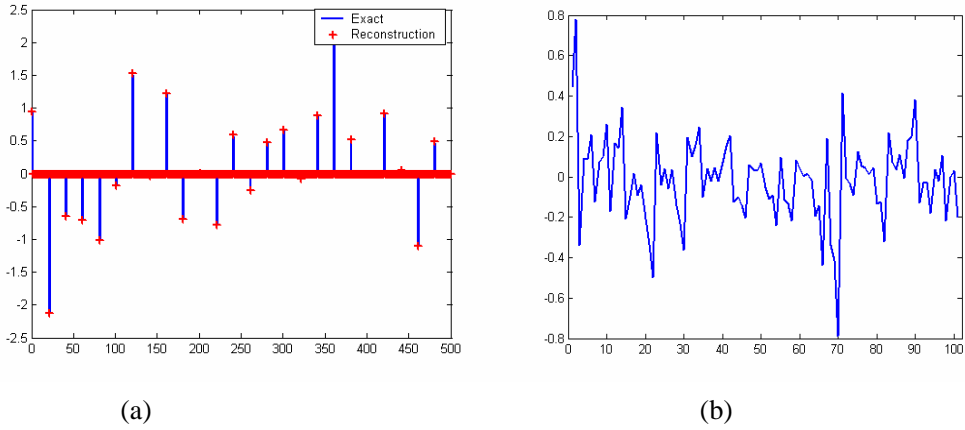

(a)                                                              (b)

Fig.4. (a) the 26-sparse reconstruction result and (b) the sampled signal.

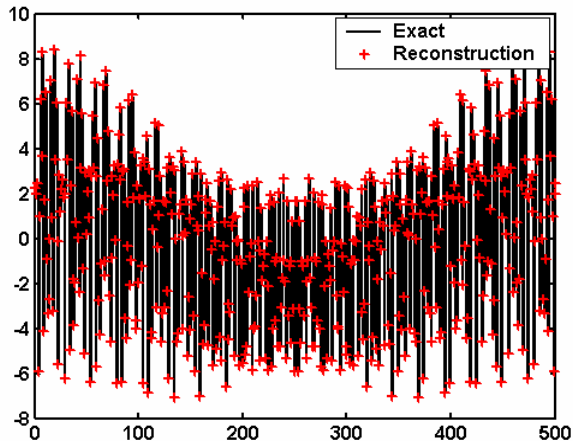

Fig.5  the reconstruction of 26-sparse signal in frequency domain

## IV. CONCLUSION AND DISCUSSION

In this paper, the design of *universal* compressive sensing filter based on *normal* lowpass (or highpass, bandpass, bandstop) filters has been developed to enable signal acquisition beyond Nyquist sampling constraints. As an example, the microstrip filter based on defected ground structure (DGS) proposed by J. I. Park et al has been employed to realize the compressive sensing filter. Of course, the general compressive sensing filter can be constructed along the identical idea by many other structures, the plasma with different electron density corresponding to different critical frequency (see Fig.6). As a matter of fact, the ionosphere can be looked as the natural compressive sensing measurement system. Some primary results are provided, where we can successfully recover signals sampled at sub-Nyquist sampling rates by exploiting additional structure other than band-limitedness. Consistent with the bonds estimated by theoretical analysis, the arbitrary *S*-sparse *n* dimensional signal can be exactly reconstructed with measurements on the order of $S\log(n)$ with overwhelming probability. Still, much further work are under investigation and will be reported in the near future, for example, the choice of optimal filter unit, the number of filters, stricter theoretical analysis about the bounds of measurements than done by Romberg [9], and so on.

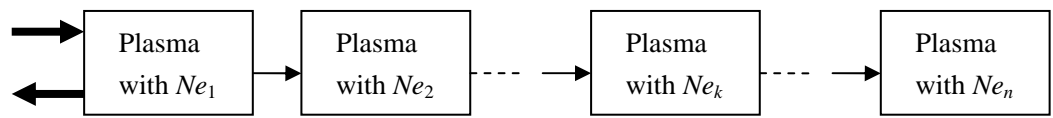

Fig. 6. the compressive sensing filter designed by using the plasma with different electron density, i.e. $Ne_1 < Ne_2 < \cdots < Ne_n$.

ACKNOWLDGEMENT:

This work has been supported by the National Natural Science Foundation of China under Grants 60701010 and 40774093.